# Bulk electronic structure of the antiferromagnetic superconducting phase in $ErNi_2B_2C$

T. Baba,[1,2] T. Yokoya,[3] S. Tsuda,[1,4] T. Kiss,[1] T. Shimojima,[1] K. Ishizaka,[1] H. Takeya,[4] K. Hirata,[4] T. Watanabe,[5] M. Nohara,[6] H. Takagi,[6] N. Nakai,[3] K. Machida,[3] T. Togashi,[2] S. Watanabe,[1] X.-Y. Wang,[7] C. T. Chen,[7] and S. Shin,[1,2]

[1]*Institute for Solid State Physics (ISSP), University of Tokyo, Kashiwa, Chiba 277-8581, Japan*

[2]*The Institute of Physical and Chemical Research (RIKEN), Sayo-gun, Hyogo 679-5143, Japan*

[3]*The Graduate School of Natural Science and Technology, Okayama University, Okayama, Okayama 700-8530, Japan*

[4]*National Institute for Material Research, Tsukuba, Ibaraki 305-0047, Japan*

[5]*Department of Physics, College of Science and Technology, Nihon University, Chiyoda-ku, Tokyo 101-8308, Japan*

[6]*Department of Advanced Materials Science, University of Tokyo, Kashiwa 277 8581, Japan*

[7]*Beijing Center for Crystal R&D, Chinese Academy of Science, Zhongguancun, Beijing, 100080, China*

## Abstract

We have performed temperature ($T$) - dependent laser-photoemission spectroscopy of antiferromagnetic (AF) superconductor $ErNi_2B_2C$ to study the electronic-structure evolution reflecting the interplay between antiferromagnetism and superconductivity. The spectra at the superconducting (SC) phase show a very broad spectral shape. $T$-dependent SC gap shows a sudden deviation from the BCS prediction just below $T_N$. This observation can be well explained by the theoretical model and thus represents characteristic bulk electronic structure of the AF SC phase for the first time.

The competition between superconductivity and magnetism has been one of the most profound and interesting problems since the publication of Ginzburg's pioneer works [1]. Chevrel phase compounds and rhodium tetra-borides are the first antiferromagnetic (AF) superconductors, where superconductivity and long-range AF order coexist, and motivated further experimental and theoretical studies [2]. The upper critical field $H_{c2}$ anomaly near the Néel temperature $T_N$ observed experimentally in the Chevrel phase compounds [3] was explained by a theory by Machida *et al*. for AF superconductors in which a spin density wave (SDW) ordering with a wave vector $Q$ may coexist with superconductivity [4]. The theory suggests that the superconducting (SC) gap $\Delta$ suddenly drops below $T_N$ due to the competition between the rapid evolution of AF molecular field and the increase of SC condensation energy. However, because of their very low transition temperatures ($T_N < 1.6$ K), no direct experimental observation of the electronic structure of the AF SC phase has been made yet.

The discovery of AF SC borocarbide compounds $R\mathrm{Ni_2B_2C}$ ($R$ = Dy, Ho, Er, and Tm) in 1994 [5,6] has activated the field for their relativity high transition $T$. Especially, $\mathrm{ErNi_2B_2C}$ has higher SC transition $T$ ($T_c$) ~ 11 K and $T_N$ ~ 6 K. Below $T_N$, $\mathrm{Er^{3+}}$ magnetic moments order in a transversely polarized incommensurate sinusoidal SDW structure with a wave vector $Q$ ~ (0.55,0,0) [7,8]. This wave vector is consistent with the predicted nesting vector, where the calculated generalized susceptibility $\chi(q)$ for $\mathrm{LuNi_2B_2C}$ shows a pronounced peak [9]. The effect of magnetic order for superconductivity has been observed in $\mathrm{ErNi_2B_2C}$ near $T_N$ in several macroscopic experiments [10-12]. In addition, scanning tunneling spectroscopy (STS) has been performed to study the electronic structures of the AF SC phase of $\mathrm{ErNi_2B_2C}$ by several groups [13,14]. Most probably due to their surface sensitivity, however, the obtained results were not consistent with each other, leaving the need of further experimental studies for the magnetic ordering effect on superconductivity.

In this letter, we report first laser-photoemission spectroscopy (PES) results of AF SC $\mathrm{ErNi_2B_2C}$ and compare with that of nonmagnetic anisotropic $s$-wave borocarbide superconductor $\mathrm{YNi_2B_2C}$ and that of $\mathrm{Y(Ni_{0.8}Pt_{0.2})_2B_2C}$ where the SC gap anisotropy is smeared by impurity scattering [15-19]. Laser-PES enables us to measure the electronic structures near the Fermi level ($E_F$) with a sub-meV energy resolution down to 2.9 K [20], with sufficient bulk sensitivity [21] due to low energy photons used. Observed $\mathrm{ErNi_2B_2C}$ spectra at the SC phase exhibit a very broad coherent peak. This is in sharp contrast to the pronounced quasiparticle peak observed in the laser-PES spectrum of $Y(Ni_{1-x}Pt_x)_2B_2C$. The $T$-dependence of the SC gap exhibits a sudden deviation from the BCS prediction just below $T_N$. This behavior is well reproduced by the Machida's theory. These observations represent characteristic electronic structure of the AF SC phase and provide deeper understanding to the competition between AF and superconductivity.

Single crystals of $ErNi_2B_2C$ and $Y(Ni_{1-x}Pt_x)_2B_2C$ were prepared with a floating zone method [22]. The dc susceptibility measurements confirmed $T_c$ = 9.3 K (midpoint) and $T_N$ = 6.0 K for $ErNi_2B_2C$ and $T_c$ = 15.4 K ($x$ = 0.0) and $T_c$ = 12.1 K ($x$ = 0.2) for $Y(Ni_{1-x}Pt_x)_2B_2C$. The estimated residual resistivity ratio (RRR) from resistivity measurements are 9.3 ($ErNi_2B_2C$), 37.4 ($YNi_2B_2C$) and 2.6 ($Y(Ni_{0.8}Pt_{0.2})_2B_2C$).

Laser-PES measurements were performed on a spectrometer built using a GAMMADATA-SCIENTA R4000WAL electron analyzer and an ultraviolet laser (hν = 6.994 eV) system [20]. The energy resolution was set to 800 μeV. Samples are cooled using a flow-type He liquid refrigerator with improved thermal shielding. The sample $T$ was measured using a silicon-diode sensor mounted below the samples and was controlled within an accuracy of 0.1 K. The vacuum of the spectrometer was better than 2 x $10^{-11}$ Torr. All the PES measurements were done for *in-situ* fractured surfaces. $T$-dependent spectral changes were confirmed by cycling $T$ across $T_c$. $E_F$ of samples was referred to that of a gold film evaporated on the sample substrate. Repeated measurements of gold $E_F$ during the present study confirmed the energy drift of ± 0.05 meV, which determined the energy accuracy of data points.

Figures 1 (a) and (b) show comparisons of the laser-PES spectra in the vicinity of $E_F$ at the same $T$ / $T_c$ ratio between $ErNi_2B_2C$ (open circles connected with a line) and $Y(Ni_{1-x}Pt_x)_2B_2C$ (filled circles connected with a line) ((a) $x$ = 0.0 and (b) $x$ = 0.2). Please note that $ErNi_2B_2C$ is in the paramagnetic phase. The laser-PES spectrum of $YNi_2B_2C$ shown in Fig. 1 (a) has two quasiparticle peaks below and above $E_F$ with a dip at $E_F$. The peak above $E_F$ is the thermally excited electrons across the SC gap at a relatively high measured $T$. On the other hand, the observed spectrum of $ErNi_2B_2C$ is very broad and the intensity at $E_F$ is relatively high. $ErNi_2B_2C$ is considered to have the same origin for superconductivity as $RNi_2B_2C$ [23]. The dramatic difference in the spectral shape of $ErNi_2B_2C$ with $YNi_2B_2C$ suggests that the broadened spectral shape of $ErNi_2B_2C$ is not explained by the SC gap anisotropy alone. In addition, the unconventional spectral shape is not entirely due to simple impurity scattering. Because $Y(Ni_{0.8}Pt_{0.2})_2B_2C$, which has a lower RRR than that of $ErNi_2B_2C$, has a sharper spectral shape than that of $ErNi_2B_2C$ (Fig. 1(b)). According to the Abrikosov-Gor'kov (AG) theory for SC alloys [24, 25], magnetic impurity scattering induces a finite lifetime $\tau_s$ of a Cooper pair and therefore this finite lifetime leads to an energy spread of $\hbar/\tau_s$. Though $ErNi_2B_2C$ is an AF superconductor, the paramagnetic phase of $ErNi_2B_2C$ having a weak exchange interaction can be accounted for by the AG theory, as discussed for Chevrel compounds [26]. So, the origin of this anomalous spectral shape might be mainly due to magnetic pair breaking effects.

In Fig. 2 (a), we show the $T$-dependent laser-PES spectra of $ErNi_2B_2C$ measured from 4.2 K (AF SC phase) to 12 K (normal phase) across $T_c$ of 9.3 K and $T_N$ of 6.0 K. The normalization of the spectra was done with the area under the curve from 20 meV to –10 meV. At the normal phase of 12 K, the spectrum has a Fermi edge whose midpoint is located at $E_F$. With decreasing $T$, the spectral edge shifts to higher binding energy with a piling up of a small quasiparticle peak. To see the difference in $T$-dependence of spectra between above and below $T_N$, we show the energy-enlarged normalized spectra of $ErNi_2B_2C$ at $T_N < T < T_c$ and $T < T_N$ in Figs. 2 (b) and 2 (c), respectively. In $T_N < T < T_c$, the PES intensity at $E_F$ gradually decreases, giving rise to a systematic shift of the leading edge and a gradual increase in the PES intensity at higher binding energy. However, at $T < T_N$, the intensity of DOS at $E_F$ hardly changes with decreasing $T$. This is not due to the drift of $E_F$, because we have checked $E_F$ just before and after the $T$-dependent measurements and found the shift of $E_F$ was within 0.1 meV. This anomalous $T$-dependence of laser-PES spectra of AF SC $ErNi_2B_2C$ is quite different from nonmagnetic superconductors, showing a characteristic electronic-structure evolution of $ErNi_2B_2C$.

To see how the SC gap behaves as a function of $T$, we fit the spectra by Dynes function using a simple isotropic $s$-wave SC gap $\Delta$ and the phenomenological broadening parameter $\Gamma$ [27] defined by $D(E,\Delta,\Gamma) = \mathrm{Re}\left\{(E - i\Gamma)\big/\sqrt{(E - i\Gamma)^2 - \Delta^2}\right\}$ for simplicity [28]. The fitting results are shown in Figs. 3 (a)-(j) as solid curves. We could fit all the $T$-dependent spectra reasonably well with value sets ($\Delta$, $\Gamma$) shown in each panel, which are defined as minimum variation of the residual [29]. We note that $\Gamma$ is larger than that of nonmagnetic SC borocarbides. This is mainly due to the magnetic pair breaking effects, as discussed above.

In Fig. 4, we plotted obtained *T*-dependent Δ and Γ values with open and filled circle, respectively. The red curve is the theoretical *T*-dependence of the SC gap [30]. Above $T_N$, obtained Δ values follow the simple BCS prediction with the SC gap magnitude of $T = 0$ ($\Delta_0$) of 1.40 meV very well. By using this value, we obtain the reduced gap value $2\Delta_0 / k_B T_c$ = 3.49, which is comparable with the mean field BCS value of 3.52. On the other hand, below $T_N$, the Δ values deviate from the BCS prediction. Figs. 3 (l) and (m) show the AF SC spectrum at 4.2K (red circles) with a simulation result (blue line) using Δ from the BCS prediction at 4.2 K and using two Γ values for fitting the peak region (l) and the edge region (m). One can easily find the simulated BCS results cannot reproduce the observed spectrum. This indicates that unusual *T*-dependence of Δ is intrinsic. Thus, we directly observed the effect of magnetic order *i.e.* a sudden reduction of Δ just below $T_N$.

To confirm whether the experimental *T*-dependence of Δ observed in the present study for $ErNi_2B_2C$ can be quantitatively reproduced by the Machida's theory, we analyzed the experimental data with the Machida's *T*-dependence of Δ defined by $\Delta(T) = gN(0)[1-\alpha m(T)]\,\pi T \sum_{\omega_n} \left( \Delta(T) \big/ \sqrt{\omega_n{}^2 + \Delta^2(T)} \right)$, where *g* is the attractive interaction constant, $m(T)$ is the normalized sublattice magnetization of the AF state, and α is determined by the following equation, $\alpha = (1/4)\pi(g_J - 1)[J(J+1)]^{1/2}(|I|/E_F gN(0))$. As shown in Fig. 4, a fitting result with $J$ = 15/2 and α = 0.235 (yellow dotted curve) well reproduces experimental data. While the *T*-dependence of Δ is reproduced by the Machida's theory, the structure due to SDW gap formation is not clearly resolved in the spectra of the AF SC phase. Thus, we also try to fit the spectra at 4.2 K using the weighted sum ($D_{SC + SDW}$) of two Dynes functions for a SC gap ($D_{SC}$) and a SDW gap ($D_{SDW}$), $D_{SC + SDW} = (1-\alpha m(T))\, D_{SC} + \alpha m(T)\, D_{SDW}$ [31]. As shown in Fig. 3 (n), the observed spectrum was found to be explained by the two Dynes functions, suggesting that the SDW gap may not be clearly observed because of large smearing effects. In addition, the estimated value $I$ = 8.7 meV from the definition of α using the value of $g_J$ = 6/5, $N(0)$ = 4.3 state/eV [32] and $gN(0)$ = 0.2 assuming that $ErNi_2B_2C$ is a weak coupling superconductor is in agreement with the value of 13 meV estimated from $d\, T_c / d$ DG (de Gennes factor) using by the AG theory [33]. Thus, these analyses indicate that the origin of the observed anomaly in *T*-dependence of Δ is a result of competition between the rapid evolution of AF molecular field and the increase of SC condensation energy with decreasing *T*.

The $T$-dependent $\Delta$ observed in $ErNi_2B_2C$ could be explained by the theoretical models on superconductivity coexisting with antiferromagnetism. We attribute this success to not only the higher energy resolution but also the bulk sensitivity of the present PES study. The broad spectral shape with extremely high zero bias conductance of the SC gap of $ErNi_2B_2C$ has been observed from STS [13,14], but the origin of the broad spectral shape could not be fully understood. In addition to the broad spectral shape with the large zero bias conductance of $ErNi_2B_2C$, $T$ smearing effects inherent in STS seem to make it difficult to do a precise determination of $T$-dependent SC gap values. On the other hand, the fact that the anomalous $T$-dependent gap observed in the present laser-PES results can be well interpreted by Machida's theory means that the energy gaps of SDW opens on the part of FSs where a nesting vector can relate two Fermi momenta. To address SC gap anisotropy as well as the relation between the SC and SDW gap in $ErNi_2B_2C$, angle resolved PES is essential and will motivate further quantitative study.

In conclusion, we have performed laser-PES on $ErNi_2B_2C$ and $Y(Ni_{1-x}Pt_x)_2B_2C$ ($x$ = 0.0 and 0.2) to investigate the electronic structure of the AF SC phase. In contrast to nonmagnetic superconductor $Y(Ni_{1-x}Pt_x)_2B_2C$ ($x$ = 0.0 and 0.2), the observed spectra of $ErNi_2B_2C$ show the very broad peak. This behavior cannot be simply explained by the SC gap anisotropy, nor simple impurity scattering. From the $T$-dependence of the SC gap, we find the sudden deviation from the BCS prediction just below $T_N$. This behavior can be well reproduced by the Machida's model. Thus the present laser-PES results represent characteristic electronic structures of AF superconductor $ErNi_2B_2C$ and lead to deeper understanding to the superconductivity coexisting with antiferomagnetism.

We thank F. Kanetaka and T. Mizokami for their help in laser-PES measurements. This study was supported by Grant-in-aid from the Ministry of Education, Science, and Culture of Japan and the Sasakawa Scientific Research Grant from The Japan Science Society.

---

Figure captions

FIG. 1: (color). Laser-PES spectra of AF SC $ErNi_2B_2C$ (open circles) and nonmagnetic $Y(Ni_{1-x}Pt_x)_2B_2C$ (filled circles connected with a line) of the same $T$ / $T_c$ ratio. (a) $x = 0.0$ and (b) $x = 0.2$.

FIG. 2: (color). (a) $T$-dependent laser-PES spectra of AF SC $ErNi_2B_2C$. Energy-enlarged spectra of (a) for two temperature regions (b) $T_N < T < T_c$ and (c) $T < T_N$.

FIG. 3: (color). (a)-(j) Results of fittings (solid lines) with $T$-dependent experimental data (open circles). (k)-(o) Energy enlargement of the AF SC spectrum at 4.2K (open circles, the size of the circles corresponds to an accuracy of data), with fitting results (solid line) using single Dynes function(k)-(m) and two Dynes functions (n)-(o). Bars at $x$ axis show the difference between the fitting results and the experimental data. See text for detail.

FIG. 4: (color). $T$-dependence of the SC gap $\Delta$ and the broadening parameter $\Gamma$ obtained from the isotropic $s$-wave Dynes function [27] analyses. Open and filled circles represent the values of Δ and Γ. The red solid and dotted lines show the predicted *T*-dependence of SC gap from the BCS theory [30] for $\Delta_0$=1.40 meV. The yellow dotted line shows the predicted *T*-dependence of SC gap from Machida's theory [4] for *J* = 15/2 and α = 0.235.

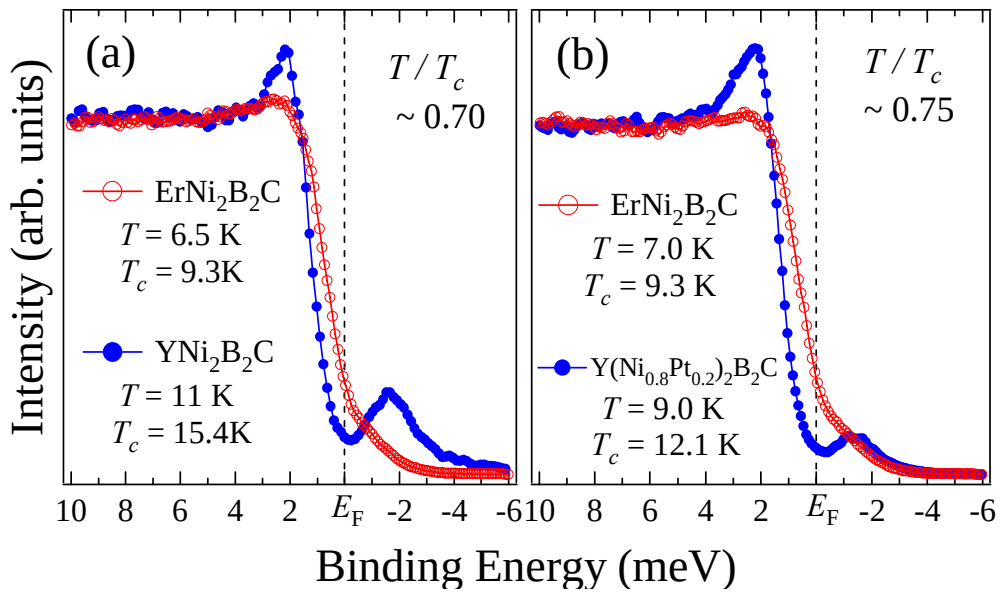

# Figure 1 (T. Baba et al.)

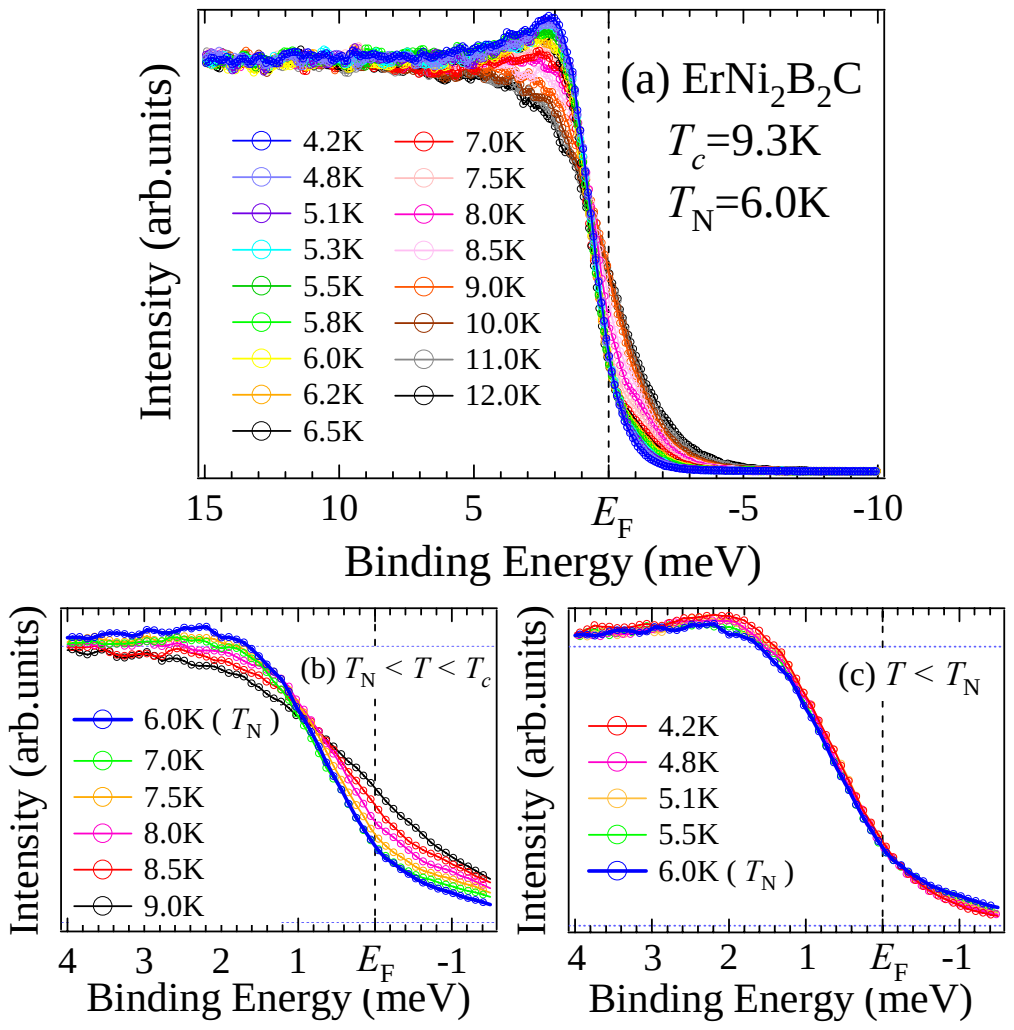

# Figure 2 (T. Baba et al.)

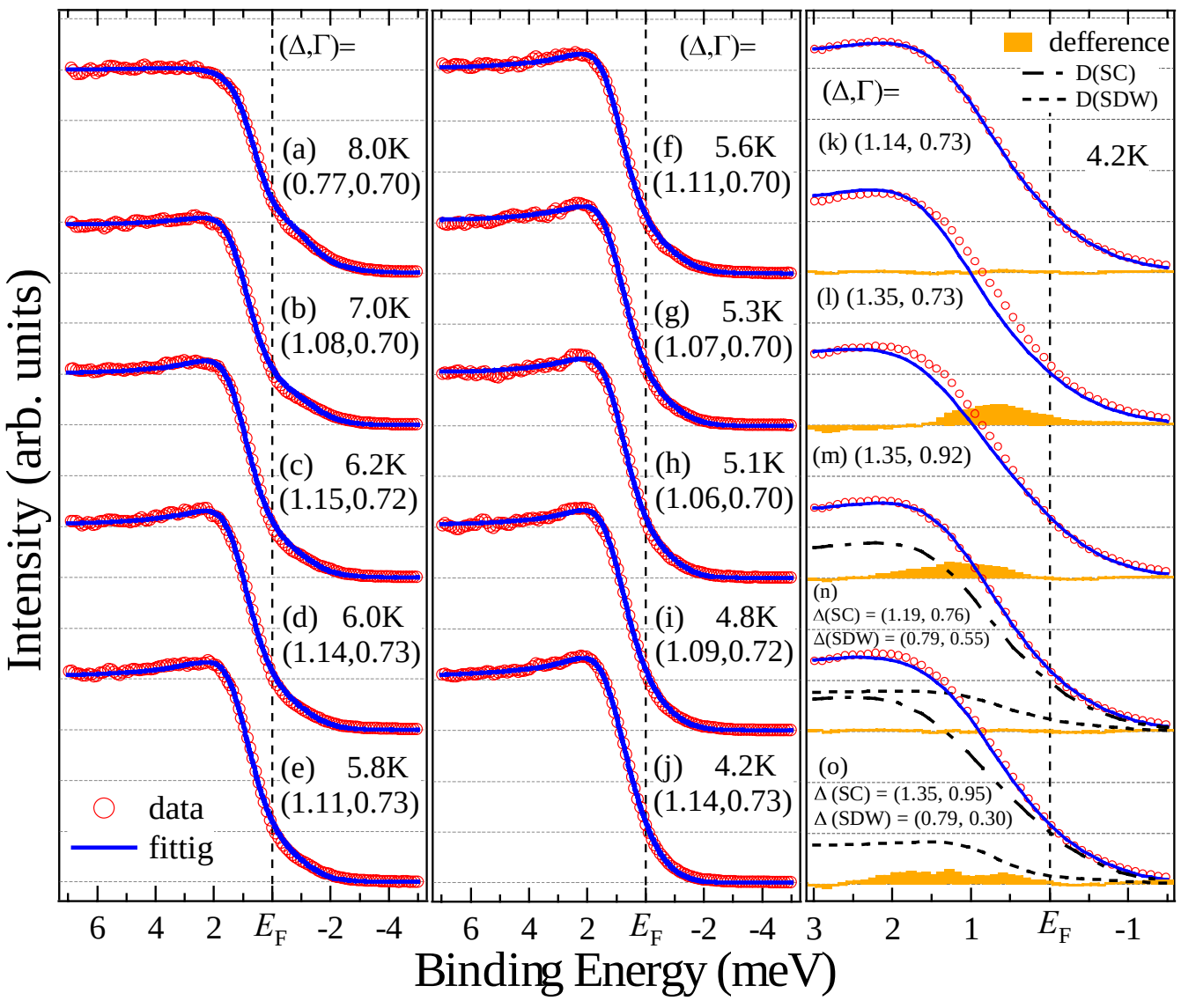

Figure 3 (T. Baba et al.)

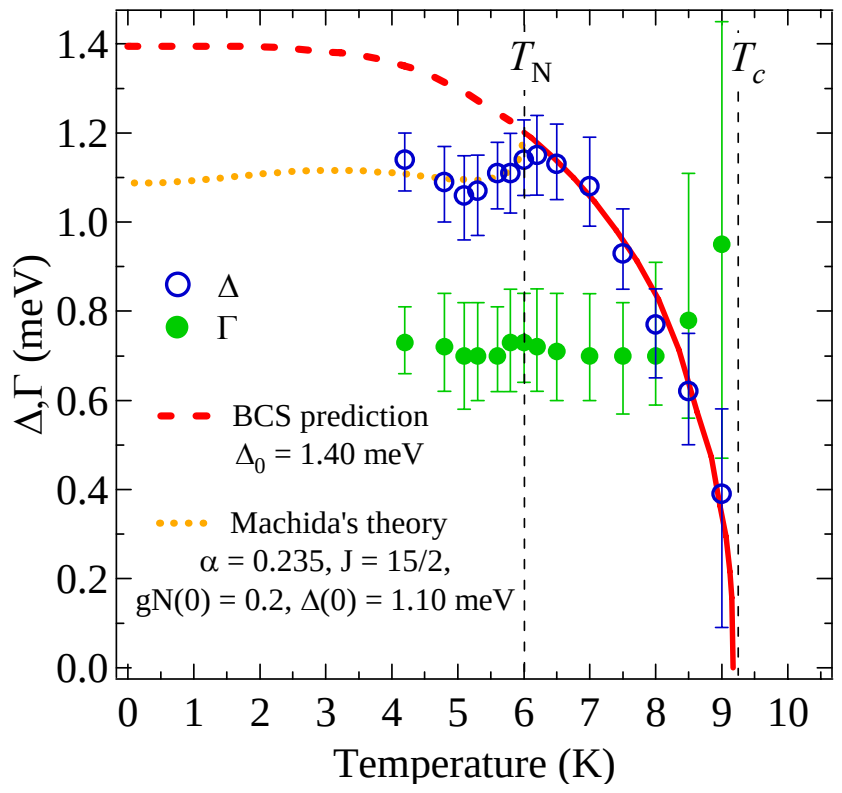

Figure 4 (T. Baba et al.)